\documentclass[reprint,a4paper,11pt]{article}

\usepackage[affil-it]{authblk}

\usepackage{amsmath,amssymb,bm}
\usepackage{graphicx}
\usepackage{xcolor}
\usepackage[colorlinks]{hyperref}
\hypersetup{
    colorlinks=true,
    linkcolor=blue,
    filecolor=magenta,
    urlcolor=[rgb]{0.00,0.00,0.50},
    citecolor=red,
    }
\usepackage[top=2cm, bottom=2cm, left=3cm, right=3cm]{geometry}
\usepackage{siunitx}
\usepackage{comment}

\usepackage[backend=bibtex,sorting=none,style=trad-abbrv,doi=true,url=false]{biblatex}
\newcommand{\Escale}{\mathcal{E}_{0}}
\newcommand{\ellL}{\ell}

\providecommand{\keywords}[1]
{
  \small
  \textbf{\textit{Keywords---}} #1
}

\newcommand{\CECs}{Centro de Estudios Cient\'{\i}ficos (CECs), Casilla 1469, Valdivia, Chile}
\newcommand{\USS}{Universidad San Sebasti\'{a}n, sede Valdivia, General Lagos 1163, Valdivia 5110693, Chile}
\newcommand{\UACh}{Instituto de Ciencias F\'isicas y Matem\'aticas, Universidad Austral  de Chile, Casilla 567, Valdivia, Chile}
\newcommand{\UCharles}{IPNP - Faculty of Mathematics and Physics, Charles University, V Hole\v{s}ovi\v{c}k\'ach 2, 18000 Prague 8, Czech Republic}
\newcommand{\INFN}{INFN, Sezione di Napoli, Complesso Universitario di Monte S. Angelo,
Via Cintia Edificio 6, 80126 Naples, Italia}
\newcommand{\UNapoli}{Dipartimento di Matematica e Applicazioni “R. Caccioppoli,” Universitá di Napoli Federico II,
Complesso Universitario di Monte S. Angelo, Via Cintia Edificio 6, 80126 Naples, Italia}

\begin{document}

\title{Casimir Sign Reversal from Finite-Range Bogoliubov Fluctuations}

\author[1,2]{Fabrizio Canfora\thanks{fabrizio.canfora@uss.cl}}
\author[3,4]{Pablo Pais\thanks{pablo.pais@matfyz.cuni.cz}}
\author[5,6]{Luigi Rosa\thanks{rosa@na.infn.it}}

\affil[1]{\USS}
\affil[2]{\CECs}
\affil[3]{\UACh}
\affil[4]{\UCharles}
\affil[5]{\INFN}
\affil[6]{\UNapoli}

\date{}

\maketitle

\begin{abstract}
We study the zero-temperature Casimir effect of a three-dimensional Bose fluid with finite-range corrections to its Bogoliubov spectrum. Beyond the universal acoustic contribution scaling as $L^{-3}$, the long-distance expansion contains a nonuniversal $L^{-5}$ correction that can compete with the phononic term. We derive this crossover analytically within a minimal isotropic finite-range model and test the asymptotic result against the full dispersion. A regulator-independent Poisson-resummed calculation confirms that sign reversals of the Casimir energy are possible, while showing that the apparent low-coupling zero inferred by balancing only the $L^{-3}$ and $L^{-5}$ terms is not quantitatively controlled. Robust sign changes instead occur in the stable regime close to roton softening, where the Casimir energy acquires an oscillatory finite-momentum contribution. Strongly dipolar atomic gases provide a natural motivation for microscopic interaction lengths in the nanometric regime. We further discuss how the resulting nonlinear force law could be probed through the frequency response of a driven nanomechanical resonator.
\end{abstract}


\newpage

\section{Introduction}

Fluctuation-induced forces provide a direct probe of the spectrum of confined quantum fields. In the electromagnetic Casimir effect the force originates from the modification of vacuum photon modes by material boundaries \cite{Casimir1948,Lifshitz1956}. In a quantum fluid the relevant fluctuating degrees of freedom are instead collective excitations. For a weakly interacting Bose gas at zero temperature the low-energy modes are phonons, and the leading Casimir energy is therefore the scalar-field result with the speed of light replaced by the sound velocity \cite{PitaevskiiStringari,Edery2006,Schiefele2009}. This acoustic approximation, however, misses the finite-momentum structure of the Bogoliubov spectrum.

Casimir energies in dilute Bose gases have previously been analyzed using the full Bogoliubov dispersion, including the crossover from the interacting phononic regime toward the ideal-gas limit \cite{Schiefele2009}. More recently, the fluctuation contribution to the surface energy of a Bose gas near hard and soft boundaries was studied within a spatially resolved Bogoliubov treatment \cite{Henkel2025}. Oscillatory Casimir interactions have also been predicted between impurities immersed in a spin--orbit-coupled Bose condensate, where the finite-momentum scale is imposed by the spin--orbit coupling and the interaction can acquire both positive and negative parts \cite{HeSunJi2017}.  The mechanism considered here is distinct: we study the zero-temperature boundary-induced Casimir energy of confined Bogoliubov modes, with the finite-momentum structure generated by a finite-range interaction and, close to instability, by the roton momentum.

Finite-range and nonlocal interactions modify the dispersion away from the strictly linear regime. They can generate curvature corrections, maxon-roton features, and, in more complete anisotropic models, supersolid behavior with vortices \cite{Casotti2024}. Such effects are especially relevant for strongly magnetic atoms, where dipolar interactions coexist with short-range scattering. Dysprosium is a natural example: measured $s$-wave scattering lengths are of order $a_s\sim 5\,\mathrm{nm}$, and the dipolar length is also of order a few nanometers \cite{Tang2015,Lahaye2009}. These microscopic lengths are much smaller than the soft-core radius of Rydberg-dressed gases, which is typically micrometric, and are therefore more compatible with nanomechanical gaps.
Here $R_{c}$ should be understood as an effective microscopic range characterizing finite-momentum corrections to the interaction, rather than as a literal Rydberg blockade radius. For dimensional estimates we consider $R_{c}$ in the nanometric regime characteristic of microscopic interaction scales in strongly dipolar gases; no quantitative identification of $R_{c}$ with either the scattering or dipolar length is implied.

Although soft-core supersolids provide a natural theoretical realization of finite-range interactions \cite{Henkel2010,Cinti2014,Ancilotto2013}, their characteristic soft-core radius is typically micrometric in Rydberg-dressed gases. They are therefore not directly compatible with the nanometric cavity separations envisioned below. In the present work we instead interpret $R_c$ as a microscopic effective range relevant to strongly dipolar atoms, while leaving the genuine soft-core supersolid regime for future work.

In this Letter we compute the zero-temperature Casimir energy of a homogeneous Bose fluid with a finite-range Bogoliubov spectrum. We deliberately use a minimal isotropic model, not a microscopic theory of a dipolar supersolid. Its purpose is to isolate analytically the effect of a finite interaction length on the confined quasiparticle zero-point energy. The full anisotropic dipolar problem, including roton softening and supersolid order, is the natural extension discussed at the end.
We also compare the long-wavelength expansion with a numerical evaluation of the full spectrum. The purpose of this comparison is not to replace the analytical result, but to determine its range of validity and to identify finite-momentum contributions that cannot be captured by a small-$q$ truncation.

We consider a fluid confined between two parallel plates of area $S$ and separation $L$, with $S\gg L^2$. For standing-wave boundary conditions in the confined direction,
\begin{equation}
q^2=q_{\parallel}^{2}+\left(\frac{n\pi}{L}\right)^2 .
\label{eq:qz}
\end{equation}
 Equation (\ref{eq:qz}) corresponds to standing-wave quantization. For periodic boundary conditions one should replace $n\pi/L$ by $2\pi n/L$, $n\in\mathbb Z$, which rescales the coefficients of the $L^{-3}$ and $L^{-5}$ terms. Below we shall also provide the expression of the Casimir energy for this case, see Eq.~(\ref{eq:main2})
The finite-range part of the interaction is represented by a repulsive soft-core potential
\begin{equation}
V(r)=V_0\,\Theta(R_c-r),
\label{eq:softcore}
\end{equation}
whose Fourier transform is
\begin{equation}
\widetilde V(q)=4\pi\, V_0\,R_c^3\,\frac{j_1(q\,R_c)}{q\,R_{c}}.
\label{eq:fourier}
\end{equation}
Here $j_1$ is the spherical Bessel function of the first type. We interpret $R_c$ as an effective microscopic finite-range scale. For a Dy-motivated estimate, one may take $R_c$ of order the scattering or dipolar length, $R_c\sim5$--$\SI{10}{\nano\meter}$.
{The top-hat interaction is used here as a minimal analytically tractable model that isolates finite-range corrections to the Bogoliubov spectrum. We do not assume a microscopic top-hat interaction for dipolar atoms. The conclusions derived below therefore concern the generic effect of finite-range dispersion rather than the detailed microscopic interaction of a particular atomic species.}

The corresponding Bogoliubov spectrum can be written as
\begin{equation}
\hbar\omega(q)=\Escale\left[A k^4+B k^2+\Lambda k j_1(k)\right]^{1/2},
\qquad k=qR_c,
\label{eq:spectrum}
\end{equation}
where
\begin{equation}
\Escale=\frac{\hbar^2}{2mR_c^2},
\qquad
\Lambda=\frac{16\pi \,m\,\rho\, V_{0}\,R_c^5}{\hbar^2} .
\label{eq:defs}
\end{equation}
Here $\rho$ is the three-dimensional particle-number density and $m$ is the mass of the bosons. The coefficient $B$ contains the acoustic contact contribution, while $A$ and $\Lambda$ encode the curvature and the finite-range part of the spectrum. Keeping $A$ and $B$ explicit allows us to separate the universal phononic contribution from the model-dependent finite-range correction.
For reference, in the contact limit $\Lambda=0$, $B=2g\rho/\Escale=(2m\,v\,R_c/\hbar)^2$ where $v$ is the sound velocity.

\section{Regularized energy}

The formal zero-point energy per unit area is
\begin{equation}
\frac{E_{0}(\ell)}{S}=\frac{\Escale}{2}\sum_n\int\frac{d^2k_{\parallel}}{(2\pi R_c)^2}
\left[A k^4+B k^2+\Lambda k j_1(k)\right]^{1/2},
\label{eq:formalE}
\end{equation}
where $k^2=k_{\parallel}^2+(n\pi/\ellL)^2$ and $\ellL=L/R_c$. The Casimir energy is the finite difference between this discrete sum and the corresponding bulk integral. Introducing an exponential regulator and subtracting the infinite-separation contribution gives
\begin{equation}
\mathcal{E}(\ell)=\lim_{\zeta\to0^+}\left[\sum_{n=0}^{\infty}{}'F_{\zeta}(n)-\int_0^{\infty}dn\,F_{\zeta}(n)\right],
\label{eq:regdef}
\end{equation}
with
\begin{equation}
F_{\zeta}(n)=\frac{\Escale}{8\pi R_c^2}
\int_{(n\pi/\ellL)^2}^{\infty}du\,
\Omega(u)e^{-\zeta u},
\label{eq:Fdef}
\end{equation}
\begin{equation}
\Omega(u)=\left[A u^2+B u+\Lambda\sqrt{u}\,j_1(\sqrt{u})\right]^{1/2}.
\label{eq:Omega}
\end{equation}

At large separations compared with the effective range, $L\gg R_c$, only the small-$u$ expansion enters the leading terms. Since
\begin{equation}
\sqrt{u}\,j_1(\sqrt{u})=\frac{u}{3}-\frac{u^2}{30}+\frac{u^3}{840}+O(u^4),
\label{eq:jexp}
\end{equation}
one obtains
\begin{equation}
\Omega(u)=\sqrt{\frac{\Lambda+3B}{3}}\,u^{1/2}
+\frac{30A-\Lambda}{20\sqrt{3(\Lambda+3B)}}\,u^{3/2}
+O(u^{5/2}).
\label{eq:Omegaexp}
\end{equation}
The Euler-Maclaurin formula then yields
\begin{equation}
\begin{aligned}
\mathcal{E}(\ell)=
-\frac{\Escale}{4R_c^2}
\Bigg[&
\frac{\pi^2\sqrt{\Lambda+3B}}{360\sqrt{3}\,\ellL^3}\\
&+\frac{\pi^4(\Lambda-30A)}{25200\sqrt{3}\,\ellL^5\sqrt{\Lambda+3B}}
+O(\ellL^{-7})
\Bigg].
\end{aligned}
\label{eq:main}
\end{equation}
This is the main result. The first term is the acoustic contribution, while the second is the leading finite-range correction. Their different powers of $L$ imply that the force law is not simply a rescaled scalar Casimir force.
{In the case of periodic boundary conditions it is easy to show that
$$ \mathcal{E}_{P}(\ell)=2\mathcal{E}(\ell/2) $$  so that
\begin{equation}
\begin{aligned}
\mathcal{E}_{P}(\ell)=
-\frac{\Escale}{4R_c^2}
\Bigg[&
16\frac{\pi^2\sqrt{\Lambda+3B}}{360\sqrt{3}\,\ellL^3}\\
&+64\frac{\pi^4(\Lambda-30A)}{25200\sqrt{3}\,\ellL^5\sqrt{\Lambda+3B}}
\Bigg].
\end{aligned}
\label{eq:main2}
\end{equation}}

Equation~(\ref{eq:main}) is an asymptotic result at fixed coupling and large $L/R_c$. We therefore test it in the Supplemental Material against a regulator-independent Poisson-resummed evaluation of the full dispersion. The comparison confirms the asymptotic expansion in its domain of validity, but also shows that a zero inferred by balancing only the displayed $L^{-3}$ and $L^{-5}$ terms need not quantitatively locate a zero of the full Casimir energy.

The acoustic limit follows by taking $A=\Lambda=0$ and $B=(2mvR_c/\hbar)^2$, giving
\begin{equation}
\mathcal E(L)=-\frac{\hbar v\pi^2}{1440L^3},
\label{eq:scalar}
\end{equation}
the scalar-field result with $c$ replaced by the sound velocity $v$. For a contact Bogoliubov spectrum, $A=1$, $\Lambda=0$, and $B=2g\rho/\Escale$, Eq.~(\ref{eq:main}) gives
\begin{equation}
\mathcal{E}_{\rm B}(\ell)=
\frac{\Escale}{4R_c^2}
\left[
\sqrt{\frac{\Escale}{2g\rho}}\frac{\pi^4}{2520\ellL^5}
-\sqrt{\frac{2g\rho}{\Escale}}\frac{\pi^2}{360\ellL^3}
\right].
\label{eq:contact}
\end{equation}
We remark that the leading part of Eq.~(\ref{eq:contact}) is equivalent to the one obtained in \cite{Edery2006}, but, in that case, finite-range contributions were not taken into account. It is for that reason that the change of sign reported in the present manuscript (see below) was found neither in \cite{Edery2006} nor in any other previous reference (at least, to the best of the authors' knowledge).


\section{Finite-range crossover and sign reversal}

The cleanest analytic illustration is obtained in the pure finite-range branch $A=1$, $B=0$. Then
\begin{equation}
\mathcal E_{\rm fr}(L)=\Escale
\left[
\frac{\pi^4(30-\Lambda)R_c^3}{100800\sqrt{3\Lambda}\,L^5}
-\frac{\pi^2\sqrt{\Lambda}R_c}{1440\sqrt{3}\,L^3}
\right].
\label{eq:finiteRangeE}
\end{equation}
The first term is generated by the leading finite-range correction to the Bogoliubov dispersion, while the second is the acoustic contribution. Within the two-term asymptotic truncation, the Casimir energy vanishes at
\begin{equation}
\Lambda_{\mathcal E=0}^{(\mathrm{as})}=\frac{30\pi^2R_c^2}{70L^2+\pi^2R_c^2},
\label{eq:EzeroLam}
\end{equation}
and the corresponding force per unit area, $\mathcal F=-\partial\mathcal E/\partial L$, vanishes at
\begin{equation}
\Lambda_{\mathcal F=0}^{(\mathrm{as})}=\frac{30\pi^2R_c^2}{42L^2+\pi^2R_c^2}.
\label{eq:FzeroLam}
\end{equation}
These equations identify where the leading acoustic and finite-range terms balance; they should not be interpreted as exact zeros of the full Casimir energy or force.

In the general case $B\neq0$, the same two-term truncation gives the candidate crossover values
\begin{equation}
\Lambda_{\mathcal E=0}^{(\mathrm{as})}=\frac{30A\pi^2-210B\ellL^2}{\pi^2+70\ellL^2}>0,
\label{eq:EzeroGeneral}
\end{equation}
for the energy and
\begin{equation}
\Lambda_{\mathcal F=0}^{(\mathrm{as})}=\frac{30A\pi^2-126B\ellL^2}{\pi^2+42\ellL^2}>0
\label{eq:FzeroGeneral}
\end{equation}
for the force. A finite acoustic/contact contribution therefore shifts the balance already at the asymptotic level. We stress here that for a realistic Dy-oriented discussion, Eqs.~(\ref{eq:EzeroGeneral}) and (\ref{eq:FzeroGeneral}) are useful as analytical diagnostics, not as quantitative phase boundaries. A full calculation using the microscopic anisotropic spectrum would ultimately be required.

The full finite-range spectrum also contains a distinct finite-momentum scale. For $A=1$ and $B=0$ it develops a zero-energy roton only at
\begin{equation}
\Lambda_c\simeq1091.45,
\qquad k_c\simeq5.4486,
\label{eq:rotoncrit}
\end{equation}

where $k_c=q_cR_c$ and the critical conditions give $5j_1(k_c)-k_cj_0(k_c)=0$ and $\Lambda_c=-k_c^3/j_1(k_c)$.

A numerical evaluation of the full Casimir sum--integral difference using Poisson resummation reveals an important qualification of the purely asymptotic picture. The apparent small-$\Lambda$ zeros of Eqs.~(\ref{eq:EzeroLam}) and (\ref{eq:FzeroLam}) occur in a nonuniform scaling regime and are not reproduced as quantitative zeros of the full energy. By contrast, robust sign changes of the full Casimir energy occur in the stable near-roton regime, before the excitation gap closes. For example, the full calculation gives

{\begin{equation}
\Lambda_{\mathcal E=0}^{\rm full}\simeq940.77(\ell=5) ,\; \Lambda_{\mathcal E=0}^{\rm full}\simeq1072.05(\ell=20)
\label{eq:fullzeros}
\end{equation}}

while $\Lambda_c\simeq1091.45$.  In the full model, the sign change therefore does not coincide exactly with the roton instability, but the finite-momentum roton sector strongly influences it. Close to the instability the Casimir energy becomes oscillatory as a function of $L$, with a period set by the roton wave number. Details and numerical convergence tests are given in the Supplemental Material.

\section{Nanomechanical detection}

The two powers in Eq.~(\ref{eq:finiteRangeE}) lead to two distinct contributions to the force. For a movable plate displaced by $x$ from the \emph{static equilibrium} gap $d$, so that $L=d+x$, it is convenient to subtract the static Casimir force already included in the definition of the equilibrium position. The equation of motion is then
\begin{equation}
\begin{aligned}
\ddot{x}+\Gamma\dot{x}+\omega_r^2x
&+\mu_4\left[\frac{1}{(d+x)^4}-\frac{1}{d^4}\right]\\
&+\mu_6\left[\frac{1}{(d+x)^6}-\frac{1}{d^6}\right]
=\frac{F_0}{m_{\rm eff}}\cos(\Omega t),
\end{aligned}
\label{eq:eom}
\end{equation}
Here $d$ denotes the static equilibrium separation including the Casimir-induced shift, where $m_{\rm eff}$ is the effective mass of the nanomechanical resonator. With the parameters used below this shift is negligible, \( |x|/d\lesssim 10^{-13}\), so that \(d\) is numerically
indistinguishable from the bare gap, and
\begin{equation}
\mu_4=\frac{S\Escale\pi^2R_c\sqrt{\Lambda}}{480\sqrt{3}\,m_{\rm eff}},
\qquad
\mu_6=\frac{S\Escale\pi^4R_c^3(\Lambda-30)}{20160\sqrt{3\Lambda}\,m_{\rm eff}}\;.
\label{eq:mu4mu6}
\end{equation}
The simple $L^{-4}+L^{-6}$ force law follows from the asymptotic energy only. If the experiment is operated in the near-roton regime where the full Casimir energy oscillates with $L$, the mechanical force should instead be obtained by differentiating the full Poisson-resummed energy. The present Duffing analysis should therefore be described as a long-wavelength proof of principle, not as the final quantitative near-roton response model.

This form is inspired by recent proposals and measurements that infer Casimir forces from nonlinear mechanical response in superconducting drum resonators \cite{deJong2025}. A successful implementation would be conceptually distinct from both standard electromagnetic Casimir-force measurements and critical Casimir experiments in classical fluids \cite{Hertlein2008,Gambassi2009}. It would amount to measuring a zero-temperature Casimir contribution generated by the confined Bogoliubov modes of an interacting quantum fluid. To the best of our knowledge, such a measurement has not yet been reported.

Introducing $y=x/d$, expanding around $y=0$, and rescaling time with a \emph{fixed} reference angular frequency,
\begin{equation}
\tau=\omega_{\rm ref} t,\qquad \omega_d=\frac{\Omega}{\omega_{\rm ref}},
\label{eq:refscale}
\end{equation}
the equation of motion becomes
\begin{equation}
\ddot y+2\alpha_1\dot y+\omega_0^2y+
\alpha_2y^2+\alpha_3y^3=f\cos(\omega_d\tau) ,
\label{eq:duffing}
\end{equation}
where, from this equation onward, dots denote derivatives with respect to the dimensionless time $\tau$ rather than the physical time $t$. The dimensionless coefficients are now defined with respect to the fixed scale $\omega_{\rm ref}$,
\begin{align}
\alpha_1&=\frac{\Gamma}{2\omega_{\rm ref}},~
\omega_0^2=\frac{\omega_r^2}{\omega_{\rm ref}^2}-4\mu-6\nu,~
\alpha_2=10\mu+21\nu, \label{eq:param} \\
\alpha_3&=-20\mu-56\nu,~
\mu=\frac{\mu_4}{d^5\omega_{\rm ref}^2},~
\nu=\frac{\mu_6}{d^7\omega_{\rm ref}^2},~
f=\frac{F_0}{m_{\rm eff}\,d\,\omega_{\rm ref}^2}\;.\nonumber
\end{align}

At first order the periodic solution can be written \cite{NayfehMook} as
\[
y(\tau)=a(\tau)\cos[\omega_d\tau-\gamma(\tau)]
+\frac{\alpha_2a^2}{2\omega_0^2}
\left[\frac{1}{3}\cos(2\omega_d\tau-2\gamma)-1\right].
\]
The slowly varying amplitude $a(\tau)$ and phase $\gamma(\tau)$ obey
\begin{eqnarray*}
\dot{a} &=&-\alpha_1 a+\frac{f}{2\omega_0}\sin\gamma,\\
a\dot{\gamma} &=&a\sigma-\frac{3\alpha}{8\omega_0}a^3
+\frac{f}{2\omega_0}\cos\gamma,
\end{eqnarray*}
where
\begin{equation}
\alpha=\alpha_3-\frac{10\alpha_2^2}{9\omega_0^2},
\qquad
\sigma=\omega_d-\omega_0.
\label{eq:alphaeff}
\end{equation}
The quantity $\sigma$ is therefore the dimensionless detuning between the drive and the linearized mechanical resonance, both expressed in units of the same fixed reference frequency.
The quadratic nonlinearity renormalizes the effective cubic coefficient. The primary-resonance response obeys \cite{NayfehMook}
\begin{equation}
\left[\alpha_1^2+\left(\sigma-\frac{3\alpha}{8\omega_0}a^2\right)^2\right]a^2
=\frac{f^2}{4\omega_0^2},
\label{eq:response}
\end{equation}

For reference, the corresponding physical linearized resonance frequency is
\begin{equation}
\omega_{\rm lin}^2=\omega_r^2-\frac{4\mu_4}{d^5}-\frac{6\mu_6}{d^7},
\qquad
\omega_0=\frac{\omega_{\rm lin}}{\Omega},
\label{eq:omegalin}
\end{equation}
and hence
\begin{equation}
\sigma=\omega_d-\omega_0
=\frac{\Omega-\omega_{\rm lin}}{\omega_{\rm ref}}.
\label{eq:detuningphysical}
\end{equation}

At stationary primary resonance, the amplitude equation can be solved for the detuning, yielding:
\begin{equation}
\sigma=\frac{3\alpha}{8\omega_0}a^2
\pm\left(\frac{f^2}{4\omega_0^2a^2}-\alpha_1^2\right)^{1/2}.
\label{eq:sigma}
\end{equation}
Equation~(\ref{eq:sigma}) is not a second definition of $\sigma$: Eq.~(\ref{eq:detuningphysical}) defines the physical detuning, whereas Eq.~(\ref{eq:sigma}) gives the value(s) of that detuning compatible with a stationary oscillation of amplitude $a$.
A change in the finite-range coefficient can therefore be detected through a static frequency shift, Eq.~(\ref{eq:param}), or through the softening/hardening of the nonlinear response curve.

\begin{figure}[t]
\centering
\includegraphics[width=0.95\columnwidth]{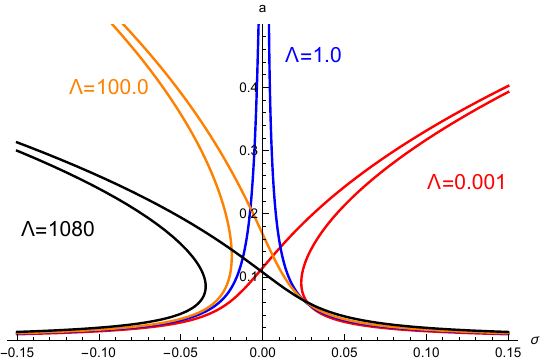}
\caption{Primary-resonance response $a(\sigma)$ obtained from Eq.~(\ref{eq:response}) for a parameter set chosen to enhance the visibility of the Casimir-induced nonlinear bending while keeping the displayed amplitude below unity. The curves show both hardening and softening behavior as $\Lambda$ is varied, with all values remaining below the finite-momentum instability threshold $\Lambda_c\simeq1091.45$; $\Lambda=1080$ probes the stable near-roton regime. The parameters are $m=\SI{2.7e-25}{\kilo\gram}$, $R_c=\SI{5e-9}{\meter}$, $d=\SI{20e-9}{\meter}$, $m_{\rm eff}=\SI{4e-14}{\kilo\gram}$, $r_{\rm plate}=\SI{9e-6}{\meter}$, $S=\pi \, r_{\rm plate}^2$, $\omega_r=2\pi\times\SI{e3}{\radian\per\second}$, $\omega_{\rm ref}=2\pi\times2\pi\times\SI{e3}{\radian\per\second}$, $\Gamma=\SI{e-10}{\radian\per\second}$, and $F_0=\SI{e-16}{\newton}$. The frequency sweep is represented by $\omega_d=\Omega/\omega_{\rm ref}$, with $\sigma=\omega_d-\omega_0$, while the coefficients are evaluated at the fixed reference scale $\omega_{\rm ref}$. These values are selected as a proof-of-principle set, not as an optimized experimental design. In particular, the extremely small damping corresponds to a very large mechanical quality factor. A realistic parameter combination should be determined by a dedicated experimental optimization. Since the perturbative derivation formally requires $a\ll1$, the upper portions of the curves should still be regarded as semi-quantitative; fully predictive response curves require direct integration of the nonlinear equation of motion, and near the roton the force derived from the full Poisson-resummed Casimir energy.}
\label{fig:response}
\end{figure}

Figure~\ref{fig:response} illustrates, within the perturbative primary-resonance description, how the finite-range Casimir contribution can change the effective nonlinear stiffness and drive the response from hardening to softening as $\Lambda$ is varied.

The parameters have been deliberately selected to make the nonlinear Casimir signature visually pronounced while keeping the motion within $|x|<d$. In particular, the reduced gap and larger plate area enhance the Casimir nonlinearity, whereas the weak drive limits the oscillation amplitude. The very small damping used in the plot corresponds to an exceptionally large mechanical quality factor and is therefore intended as an illustrative sensitivity-enhancement parameter rather than as a claim of present experimental performance. A realistic operating point should ultimately be chosen by an experimental implementation through a joint optimization of $Q$, $d$, $S/m_{\rm eff}$, $F_0$, and the near-resonant drive frequency. Importantly, all values of $\Lambda$ shown in Fig.~\ref{fig:response} lie below the finite-momentum instability threshold $\Lambda_c\simeq1091.45$; the largest value, $\Lambda=1080$, probes the stable near-roton regime without crossing the instability.

A quantitative prediction of the experimentally accessible response should instead be obtained by direct numerical integration of the full nonlinear equation of motion. At the first level of improvement one should retain the unexpanded algebraic force in Eq.~(\ref{eq:eom}), rather than its cubic Duffing truncation. In the near-roton regime, where the full Casimir energy develops an oscillatory dependence on $L$, the most faithful description is obtained by replacing the asymptotic $L^{-4}+L^{-6}$ force law by the force derived from the full Poisson-resummed energy,
\begin{equation}
F_{\rm Cas}^{\rm full}(L)=-S\,\frac{\partial \mathcal E_{\rm Cas}^{\rm full}(L)}{\partial L}.
\label{eq:fullforce}
\end{equation}
Thus the perturbative Duffing analysis, the unexpanded asymptotic force law, and the full Poisson-resummed force should be regarded as a hierarchy of increasingly quantitative descriptions of the same mechanical signature. We also emphasize that the softening--hardening crossover shown in Fig.~\ref{fig:response} is not identical to the zero of the Casimir energy or force; it is a dynamical consequence of the Casimir-induced nonlinear terms in the mechanical equation of motion.

\section{Conclusions}

We have derived the zero-temperature Casimir energy of a confined Bose fluid with finite-range corrections to the Bogoliubov spectrum (which are relevant, for instance, in the theory of supersolid). At large separation the result separates a universal acoustic contribution, proportional to $L^{-3}$, from a finite-range correction proportional to $L^{-5}$. Their competition provides a transparent analytical mechanism for modifying the sign and range dependence of the Casimir interaction. A full Poisson-resummed evaluation adds an important qualification: the zero obtained by balancing only these two asymptotic terms is not quantitatively controlled in the low-coupling scaling regime. Nevertheless, the full dispersion does support genuine sign reversals in the stable region close to roton softening, where finite-momentum modes generate an additional oscillatory contribution. Strongly dipolar atoms such as dysprosium provide a natural motivation for nanometric effective lengths, making this regime more compatible with nanomechanical geometries than Rydberg-dressed soft-core gases.

The present calculation should be viewed as a controlled analytic benchmark rather than a microscopic theory of a dipolar supersolid. The near-roton oscillations found here are qualitatively reminiscent of the oscillatory impurity-mediated Casimir potential in spin--orbit-coupled Bose condensates \cite{HeSunJi2017}, in the broad sense that a finite characteristic momentum imprints an oscillation scale on the interaction. The microscopic origin and geometry are, however, different: in the present problem the oscillation follows from boundary confinement of the full finite-range Bogoliubov spectrum and its period is set by the roton wave number.
The natural next step is to replace the isotropic effective potential by the full anisotropic dipole-dipole interaction, including the geometry-dependent roton minimum and Lee-Huang-Yang stabilization. Such an extension would determine whether the finite-range Casimir sign reversal survives in the experimentally realized dipolar supersolid regime, comparing with recent results \cite{Napiorkowski2026}.

A distinct nanostructure-compatible direction is provided by spatially indirect excitons in semiconductor double layers \cite{Joglekar2006,Conti2023,Zeng2026}. In these systems the relevant geometric length scales are naturally nanometric. Moreover, theoretical work predicts Wigner or Chester supersolid phases of indirect excitons, while recent experiments report a superfluid-to-insulator transition consistent with an ordered excitonic solid. The interaction is not of soft-core top-hat form, but rather dipolar and geometry-dependent; extending the present Casimir calculation to this setting would therefore require replacing the finite-range Bogoliubov spectrum by the appropriate excitonic dipolar dispersion.

\section*{Acknowledgments}
F.~C. has been supported by FONDECYT Grant No.~1240048 and by Grant ANID EXPLORACION No.~13250014. P.~P. gladly acknowledge support from Charles University Research Center (UNCE 24/SCI/016). L.~R. gratefully acknowledge the CECs for the kind hospitality.

\clearpage
\pagebreak

\newpage
\begin{center}
    {\Large \bfseries Supplemental Material for \\[0.5em] ``Casimir Sign Reversal from Finite-Range Bogoliubov Fluctuations''}
\end{center}
\vspace{1.5em}

\setcounter{equation}{0}
\setcounter{figure}{0}
\setcounter{table}{0}
\setcounter{section}{0}
\setcounter{page}{1}

\renewcommand{\theequation}{S\arabic{equation}}
\renewcommand{\thefigure}{S\arabic{figure}}
\renewcommand{\thetable}{S\arabic{table}}
\renewcommand{\thesection}{S\Roman{section}}
\renewcommand{\thepage}{S\arabic{page}}


In this Supplemental Material, we provide detailed derivations of the analytical results discussed in the main text.


\section{Fourier Transform of the Soft-Core Potential}
\label{sec:fourier}

We consider the finite-range repulsive soft-core potential given by
\begin{equation}
    V(r) = V_0 \Theta(R_c - r).
\end{equation}
To obtain the momentum-space representation of the interaction, we calculate the three-dimensional Fourier transform, $\tilde{V}(q) = \int d^3r e^{-i\mathbf{q}\cdot\mathbf{r}} V(r)$.
Recalling the definition of the spherical Bessel function of the first kind of order one, $j_1(x) = (\sin x - x\cos x)/x^2$, we can rewrite the expression in terms of $x = qR_c$ to obtain the form used in the main text:
\begin{equation}
    \tilde{V}(q) = 4\pi V_0 R_c^3 \frac{j_1(qR_c)}{qR_c}\;.
\end{equation}

\section{Modified Bogoliubov Spectrum}
\label{sec:spectrum}

The elementary excitation spectrum for a weakly interacting Bose gas is given by the standard Bogoliubov dispersion relation,
\begin{equation}
    \hbar\omega(q) = \sqrt{\epsilon_q \left( \epsilon_q + 2\rho\tilde{V}_{\text{eff}}(q) \right)} ,
\end{equation}
where $\epsilon_q = \hbar^2 q^2 / 2m$ is the free-particle kinetic energy ($m$ being the particle mass), $\rho$ is the bulk density, and $\tilde{V}_{\text{eff}}(q)$ is the effective interaction in momentum space.

To work with dimensionless quantities, we define the scaled momentum $k = qR_c$ and the characteristic energy scale $E_0 = \hbar^2 / 2mR_c^2$. The kinetic energy can then be expressed as $\epsilon_q = E_0 k^2$.

In our model, the interaction consists of a local contact contribution, $g=4\pi\hbar^2a_s/m$, and the finite-range soft-core potential derived in the main text. The soft-core contribution to the interaction energy is
\begin{equation}
    2\rho\tilde{V}(q)=\frac{8\pi\rho V_0R_c^3}{k}j_1(k).
\end{equation}
The contact contribution produces the acoustic term $2\rho g\,\epsilon_q$. Dividing the full squared dispersion by $E_0^2$ gives
\begin{equation}
    (\hbar\omega)^2=E_0^2\left[k^4+B k^2+\Lambda k j_1(k)\right],
\end{equation}
where
\begin{equation}
    \Lambda=\frac{8\pi\rho V_0R_c^3}{E_0}
    =\frac{16\pi m\rho V_0R_c^5}{\hbar^2},
\end{equation}
and
\begin{equation}
    B=\frac{2\rho g}{E_0}
    =\left(\frac{2mcR_c}{\hbar}\right)^2,
    \qquad c^2=\frac{\rho g}{m}.
\end{equation}
Thus the top-hat Bogoliubov model itself gives
\begin{equation}
    \hbar\omega(q)=E_0\left[k^4+B k^2+\Lambda k j_1(k)\right]^{1/2}.
\end{equation}
For comparison with the notation of the main text we may write the more general phenomenological form
\begin{equation}
    \hbar\omega(q)=E_0\left[A k^4+B k^2+\Lambda k j_1(k)\right]^{1/2},
\end{equation}
where the microscopic top-hat Bogoliubov model corresponds to $A=1$. Keeping $A$ explicit is useful for tracking the curvature contribution algebraically, but it should not be interpreted as an additional independent microscopic parameter of the strict top-hat model.

\section{Transverse Integration and Regularized Casimir Energy}
\label{sec:regularization}

To evaluate the Casimir force between two parallel plates separated by a distance $L$, we calculate the zero-point energy of the Bogoliubov modes. The confinement imposes the physical wave numbers $q_z=n\pi/L$. Since $k=qR_c$ is dimensionless, the corresponding dimensionless confined component is
\begin{equation}
    k_z=q_zR_c=\frac{n\pi}{\ell},\qquad \ell=\frac{L}{R_c}.
\end{equation}
Hence $k^2=k_\parallel^2+(n\pi/\ell)^2$, where $k_\parallel=q_\parallel R_c$ is the dimensionless in-plane momentum.

The formal zero-point energy per unit area is the sum over all allowed modes. To avoid confusing this quantity with the energy scale $E_0$, we denote it by $\mathcal E_{\rm ZP}(L)$:
\begin{equation}
    \frac{\mathcal E_{\rm ZP}(L)}{S} = \frac{E_0}{2} \sum_n \int \frac{d^2k_\parallel}{(2\pi R_c)^2} \Omega(k^2) ,
\end{equation}
where $\Omega(k^2) = [ A k^4 + B k^2 + \Lambda k j_1(k) ]^{1/2}$ is the dimensionless dispersion.

To compute this quantity, we introduce a smooth exponential regulator $e^{-\zeta k^2}$ (with $\zeta \to 0^+$) to tame the ultraviolet divergence. Passing to polar coordinates in the transverse momentum plane, $d^2k_\parallel = 2\pi k_\parallel dk_\parallel$. We introduce the change of variables $u = k^2$, which gives $du = 2 k_\parallel dk_\parallel$, and therefore $d^2k_\parallel = \pi du$.

For a fixed mode $n$, the lower limit of integration corresponds to $k_\parallel = 0$, meaning $u = (n\pi/\ell)^2$. The integral over the transverse directions for a single mode is defined as $F_\zeta(n)$,
\begin{equation}
    F_\zeta(n) = \frac{E_0}{2 (2\pi R_c)^2} \int_{(n\pi/\ell)^2}^\infty (\pi du) \, \Omega(u) e^{-\zeta u}.
\end{equation}
where
\begin{equation}
    \Omega(u) = \left[ A u^2 + B u + \Lambda \sqrt{u} j_1(\sqrt{u}) \right]^{1/2}.
\end{equation}
The physical Casimir energy is the finite difference between the discrete sum of modes in the confined geometry and the continuous integral over the bulk space, evaluated as the regulator is removed:
\begin{equation}
    E(L) = \lim_{\zeta \to 0^+} \left[ {\sum_{n=0}^\infty}^\prime F_\zeta(n) - \int_0^\infty dn \, F_\zeta(n) \right] ,
\end{equation}
where the prime on the summation indicates that the $n=0$ term is multiplied by a factor of $1/2$.

\section{Large-distance expansion and its domain of validity}
\label{sec:asymptotic}

For $u\ll1$,
\begin{equation}
\sqrt{u}\,j_1(\sqrt{u})=\frac{u}{3}-\frac{u^2}{30}+\frac{u^3}{840}+O(u^4),
\end{equation}
so that
\begin{equation}
\Omega(u)=\sqrt{\frac{\Lambda+3B}{3}}u^{1/2}
+\frac{30A-\Lambda}{20\sqrt{3(\Lambda+3B)}}u^{3/2}
+O(u^{5/2}).
\end{equation}
Applying Euler--Maclaurin to the regulated sum--integral difference yields
\begin{eqnarray}
\mathcal E_{\rm as}(L) &=&
-\frac{E_0}{4R_c^2}\left[
\frac{\pi^2\sqrt{\Lambda+3B}}{360\sqrt3\,\ell^3}
 \right.\label{eq:SMasym} \\
 &+&\left. \frac{\pi^4(\Lambda-30A)}{25200\sqrt3\,\ell^5\sqrt{\Lambda+3B}}
+O(\ell^{-7})\right]. \nonumber
\end{eqnarray}
Equation~(\ref{eq:SMasym}) is an expansion at fixed $A$, $B$, and $\Lambda$ in the limit $\ell=L/R_c\to\infty$. This point matters when the first two terms are used to infer a zero. In the $A=1$, $B=0$ branch, balancing the displayed $\ell^{-3}$ and $\ell^{-5}$ pieces gives $\Lambda\sim\ell^{-2}$ at large $\ell$. In such a correlated limit, nominally higher-order terms can become comparable to the retained terms, and the two-term zero need not approximate the zero of the full Casimir energy.

{{\bf{N.B.}} This is the mathematical reason why the analytical zero should be described as a crossover scale of the truncated asymptotic expansion rather than as an exact physical threshold.}

\section{Poisson-resummed evaluation of the full Casimir energy}
\label{sec:poisson}

To test the full spectrum without extrapolating an ultraviolet regulator, we use the half-line Poisson identity
\begin{equation}
\sum_{n=0}^{\infty}{}'F(n)-\int_0^\infty dx\,F(x)
=2\sum_{p=1}^{\infty}\int_0^\infty dx\,F(x)\cos(2\pi px).
\label{eq:poisson}
\end{equation}
Applying Eq.~(\ref{eq:poisson}) to the transverse-integrated energy and interchanging the order of integration gives a sine-transform representation. For $A>0$ we first subtract the polynomial ultraviolet pieces
\begin{equation}
R(k)=\Omega(k)-\sqrt A\,k^2-\frac{B}{2\sqrt A},
\end{equation}
which do not contribute to the finite Casimir difference under analytic regularization. Defining
\begin{equation}
\overline{\mathcal E}(\ell,\Lambda)
\equiv\frac{\mathcal E_{\rm Cas}}{E_0/(8\pi R_c^2)},
\end{equation}
we obtain the regulator-independent representation
\begin{equation}
{
\overline{\mathcal E}(\ell,\Lambda)
=\frac{2}{\pi}\sum_{p=1}^{\infty}\frac1p
\int_0^\infty dk\,kR(k)\sin(2p\ell k)
}.
\label{eq:poissonmaster}
\end{equation}
For the branch used for the analytical illustration, $A=1$, $B=0$,
\begin{equation}
R(k)=\sqrt{k^4+\Lambda k j_1(k)}-k^2.
\end{equation}
At large $k$, $kR(k)=\Lambda j_1(k)/2+O(k^{-5})$. In the numerical implementation we integrate the low- and intermediate-momentum part directly and treat the leading oscillatory ultraviolet tail analytically. The Poisson-image sum is then convergent and can be checked systematically by increasing both the momentum cutoff and the number of images.

\section{Numerical comparison: asymptotic versus full result}
\label{sec:numerical}

The Poisson calculation confirms two distinct statements.

First, at fixed coupling and sufficiently large $\ell$, the full result approaches Eq.~(\ref{eq:SMasym}), as expected. The smooth acoustic background and its leading finite-range correction are therefore correctly described by the $\ell^{-3}+\ell^{-5}$ expansion.

Second, the apparent low-coupling zero obtained by balancing only these two terms is not reproduced quantitatively by the full spectrum. The reason is the nonuniform scaling discussed in the preceding section. Thus the analytical zero is useful as an indicator of competition between the acoustic and finite-range terms, but not as an exact location of the full sign reversal.

The full calculation nevertheless displays robust zeros before the roton instability. For $A=1$ and $B=0$, the bulk excitation spectrum first becomes unstable at
\begin{equation}
\Lambda_c\simeq1091.45,\qquad k_c\simeq5.4486.
\end{equation}
Solving Eq.~(\ref{eq:poissonmaster}) for the zeros of the Casimir energy gives, for example,
\begin{align}
\Lambda_{\mathcal E=0}^{\rm full}&\simeq940.7718, && \ell=5,\\
\Lambda_{\mathcal E=0}^{\rm full}&\simeq1072.0507, && \ell=20.
\end{align}
Both values lie below $\Lambda_c$: the Casimir sign reversal therefore occurs while the homogeneous Bogoliubov spectrum is still stable. At the same time, these values are close enough to the roton region that the finite-momentum structure of the spectrum has become dynamically important.

\begin{figure}[t]
\centering
\includegraphics[width=0.98\columnwidth]{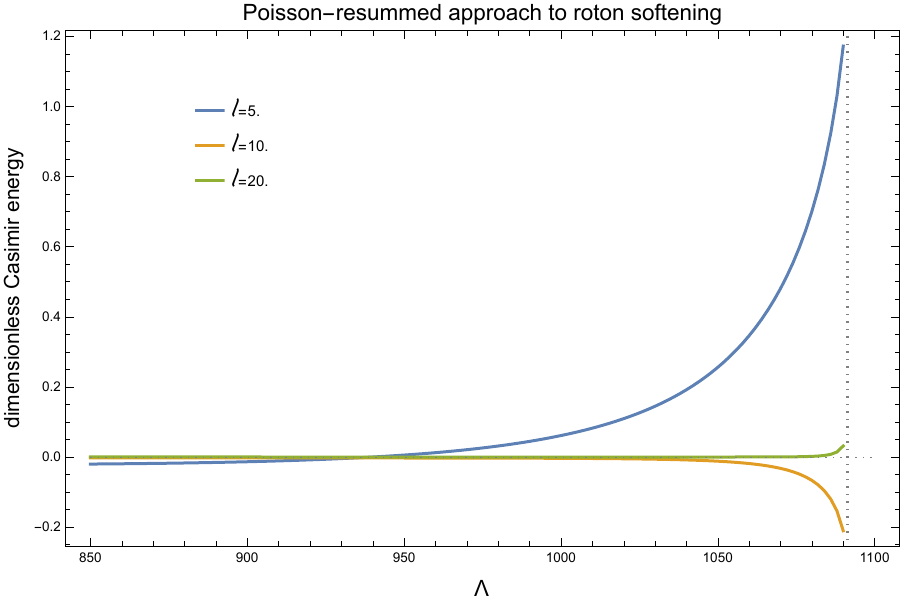}
\caption{Poisson-resummed Casimir energy obtained from the full $A=1$, $B=0$ Bogoliubov dispersion as $\Lambda$ approaches the roton instability. The horizontal zero line separates positive and negative Casimir energy. The sign depends nonmonotonically on $L/R_c$, demonstrating that the full sign reversal is controlled by finite-momentum physics not contained in the two-term small-$q$ expansion.}
\label{fig:poissonlambda}
\end{figure}

Figure~\ref{fig:poissonlambda} shows the full Poisson-resummed Casimir energy as the coupling approaches the roton threshold for several separations. The result is strongly nonmonotonic in $\ell$: different plate separations can have opposite Casimir-energy signs at the same value of $\Lambda$.

\section{Near-roton oscillations and their physical origin}
\label{sec:rotonosc}

The Poisson representation directly exposes the finite-momentum origin of the oscillatory behavior. If the non-acoustic part of $kR(k)$ is concentrated around a roton minimum $k_r$, the leading Poisson image contains a contribution of the form
\begin{equation}
\delta\overline{\mathcal E}_r(\ell)
\sim \mathcal A(\ell)\sin(2\ell k_r+\phi),
\end{equation}
where $\mathcal A(\ell)$ is a slowly varying envelope. The expected oscillation period is therefore
\begin{equation}
\Delta\ell\simeq\frac{\pi}{k_r},
\qquad
\Delta\ell_{\rm zero}\simeq\frac{\pi}{2k_r}.
\label{eq:rotonperiod}
\end{equation}
For $\Lambda=1080$, minimization of the full dispersion gives
\begin{equation}
k_r=5.4427984,
\end{equation}
so that
\begin{equation}
\frac{\pi}{k_r}=0.5772017,
\qquad
\frac{\pi}{2k_r}=0.2886009.
\end{equation}
The numerically observed mean spacing between consecutive zeros is $0.2885226$, while a Fourier analysis gives a dominant period $0.57381$. The agreement with Eq.~(\ref{eq:rotonperiod}) shows that the oscillation is tied to the physical roton momentum rather than to the numerical cutoff or grid.


\section{Implications for the interpretation of the analytical result}

The analytical and numerical calculations play complementary roles. Equation~(\ref{eq:SMasym}) cleanly isolates the universal acoustic $L^{-3}$ term and the leading finite-range $L^{-5}$ correction, and is the appropriate controlled result at fixed coupling and $L\gg R_c$. The full Poisson-resummed calculation is required when finite-momentum structure becomes important. It shows that the physical sign reversal is real, but that its quantitative location cannot in general be inferred by equating only the first two terms of the asymptotic series.


\section{Nanomechanical detection}
\label{sec:nanomech}

Within the long-wavelength approximation, differentiating the $L^{-3}$ and $L^{-5}$ terms of the asymptotic Casimir energy produces force contributions proportional to $L^{-4}$ and $L^{-6}$. We write the instantaneous separation as
\begin{equation}
L=d+x,
\end{equation}
where $d$ is the \emph{static equilibrium} gap. Since the static Casimir force is already included in the definition of that equilibrium position, the constant contribution at $x=0$ must be subtracted. The mechanical equation is therefore
\begin{equation}
\begin{aligned}
\ddot{x}+\Gamma\dot{x}+\omega_r^2x
&+\mu_4\left[\frac{1}{(d+x)^4}-\frac{1}{d^4}\right]\\
&+\mu_6\left[\frac{1}{(d+x)^6}-\frac{1}{d^6}\right]
=\frac{F_0}{m_{\rm eff}}\cos(\Omega t)\;,
\end{aligned}
\label{eq:SMfullEOM}
\end{equation}
where $m_{eff}$ is the effective mass of the resonator.
The force coefficients are
\begin{equation}
\mu_4=\frac{S E_0\pi^2R_c\sqrt{\Lambda}}
{480\sqrt3\,m_{\rm eff}},
\qquad
\mu_6=\frac{S E_0\pi^4R_c^3(\Lambda-30)}
{20160\sqrt{3\Lambda}\,m_{\rm eff}}.
\label{eq:SMmu}
\end{equation}
Both coefficients have dimensions such that $\mu_4/L^4$ and $\mu_6/L^6$ are accelerations. Equation~(\ref{eq:SMfullEOM}) is therefore dimensionally consistent term by term.

We now introduce
\begin{equation}
y=\frac{x}{d}
\end{equation}
and scale time with a \emph{fixed} reference angular frequency,
\begin{equation}
\tau=\omega_{\rm ref}t,
\qquad
\omega_d=\frac{\Omega}{\omega_{\rm ref}}.
\label{eq:SMrefscale}
\end{equation}
This choice is preferable to scaling time by the swept drive frequency itself, because all mechanical coefficients then remain fixed during a frequency scan. Expanding Eq.~(\ref{eq:SMfullEOM}) through cubic order in $y$ gives
\begin{equation}
\ddot y+2\alpha_1\dot y+\omega_0^2y
+\alpha_2y^2+\alpha_3y^3
=f\cos(\omega_d\tau),
\label{eq:SMduffing}
\end{equation}
where dots now denote derivatives with respect to $\tau$, and
\begin{align}
\alpha_1&=\frac{\Gamma}{2\omega_{\rm ref}},
&\omega_0^2&=\frac{\omega_r^2}{\omega_{\rm ref}^2}-4\mu-6\nu,
\nonumber\\
\alpha_2&=10\mu+21\nu,
&\alpha_3&=-20\mu-56\nu,
\label{eq:SMcoeffs}\\
\mu&=\frac{\mu_4}{d^5\omega_{\rm ref}^2},
&\nu&=\frac{\mu_6}{d^7\omega_{\rm ref}^2},
\qquad
f=\frac{F_0}{m_{\rm eff}\,d\,\omega_{\rm ref}^2}.
\nonumber
\end{align}
The corresponding physical linearized resonance frequency is
\begin{equation}
\omega_{\rm lin}^2=\omega_r^2-\frac{4\mu_4}{d^5}-\frac{6\mu_6}{d^7},
\qquad
\omega_0=\frac{\omega_{\rm lin}}{\omega_{\rm ref}}.
\label{eq:SMomegalin}
\end{equation}

Near primary resonance, the quadratic term renormalizes the cubic nonlinearity. Defining
\begin{equation}
\alpha_{\rm eff}=\alpha_3-
\frac{10\alpha_2^2}{9\omega_0^2},
\label{eq:SMeffalpha}
\end{equation}
and the dimensionless detuning
\begin{equation}
\sigma=\omega_d-\omega_0
=\frac{\Omega-\omega_{\rm lin}}{\omega_{\rm ref}},
\label{eq:SMdetuning}
\end{equation}
the stationary primary-resonance response is
\begin{equation}
\left[\alpha_1^2+
\left(\sigma-\frac{3\alpha_{\rm eff}}{8\omega_0}a^2\right)^2\right]a^2
=\frac{f^2}{4\omega_0^2}.
\label{eq:SMresponse}
\end{equation}
Solving for the detuning gives
\begin{equation}
\sigma(a)=\frac{3\alpha_{\rm eff}}{8\omega_0}a^2
\pm\left(\frac{f^2}{4\omega_0^2a^2}-\alpha_1^2\right)^{1/2}.
\label{eq:SMsigmaofa}
\end{equation}
Equation~(\ref{eq:SMdetuning}) is the \emph{definition} of the physical detuning, while Eq.~(\ref{eq:SMsigmaofa}) is the stationary response relation that gives the value(s) of that detuning compatible with a chosen amplitude $a$. The two equations therefore play different roles and are not competing definitions of $\sigma$.

The perturbative description formally requires $a\ll1$. Parameter sets used to illustrate the nonlinear bending should therefore be interpreted as proof-of-principle choices rather than optimized experimental operating points, especially when the upper part of a curve reaches amplitudes of order a few tenths. In addition, the long-wavelength force law itself becomes insufficient near the roton region. A quantitatively predictive treatment should first retain the unexpanded force in Eq.~(\ref{eq:SMfullEOM}), and in the near-roton regime should replace the asymptotic $L^{-4}+L^{-6}$ force by
\begin{equation}
F_{\rm Cas}^{\rm full}(L)
=-S\frac{\partial\mathcal E_{\rm Cas}^{\rm full}(L)}{\partial L},
\end{equation}
with $\mathcal E_{\rm Cas}^{\rm full}$ obtained from the Poisson-resummed calculation.

\printbibliography[title={References}]

\end{document}